\documentclass[12pt]{article}
  \usepackage{amsfonts}
  \usepackage{amsmath}
\usepackage{amssymb}
\usepackage{amscd}
\usepackage[dvips]{graphicx}
\begin{document}

~~
\bigskip
\bigskip
\begin{center}
{\Large {\bf{{{Generating of additional force terms in Newton equation by twist-deformed Hopf algebras
and   classical symmetries}}}}}
\end{center}
\bigskip
\bigskip
\bigskip
\begin{center}
{{\large ${\rm {Marcin\;Daszkiewicz}}$}}
\end{center}
\bigskip
\begin{center}
\bigskip

{ ${\rm{Institute\; of\; Theoretical\; Physics}}$}

{ ${\rm{ University\; of\; Wroclaw\; pl.\; Maxa\; Borna\; 9,\;
50-206\; Wroclaw,\; Poland}}$}

{ ${\rm{ e-mail:\; marcin@ift.uni.wroc.pl}}$}

\end{center}
\bigskip
\bigskip
\bigskip
\bigskip
\bigskip
\bigskip
\bigskip
\bigskip
\bigskip
\begin{abstract}
We compare  two ways of force terms generating in the model of nonrelativistic particle moving in the presence of constant field force $\vec{F}$.
First of them uses the twist-deformed acceleration-enlarged Newton-Hooke quantum space-times while the second one incorporates the doubly
enlarged Newton-Hooke transformations of classical space. Particulary, we find the conditions for which the both treatments provide
the same force terms.
\end{abstract}
\bigskip
\bigskip
\bigskip
\bigskip
\eject

\section{{{Introduction}}}

In the last time, there appeared a lot of papers dealing with
classical and quantum  mechanics (see e.g. \cite{mech}-\cite{qmnext})
as well as with field theoretical models (see e.g. \cite{field}), in
which  the quantum space-time  plays a crucial role.
The idea to use noncommutative coordinates is quite old - it goes back to Heisenberg
and was firstly formalized by Snyder in   \cite{snyder}.  Recently, however, there were found new formal
arguments based mainly on Quantum Gravity \cite{grav1}
and String Theory models \cite{string1},
indicating that space-time at Planck scale  should be
noncommutative, i.e. it should  have a quantum nature. Besides,
 the main reason for such considerations follows from the
suggestion  that relativistic space-time symmetries should be
modified (deformed) at Planck scale, while  the classical Poincare
invariance still remains valid at
larger distances \cite{1a}, \cite{1anext}.

Presently, it is well known, that in accordance with the
Hopf-algebraic classification of all deformations of relativistic
and nonrelativistic symmetries, one can distinguish three basic
types of quantum spaces \cite{class1}, \cite{class2}. First of them
corresponds to the well-known canonical type of noncommutativity
\begin{equation}
[\;{\hat x}_{\mu},{\hat x}_{\nu}\;] =
i\theta_{\mu\nu}\;,
\label{wielkaslawia}
\end{equation}
with antisymmetric constant tensor $\theta^{\mu\nu}$. Its
relativistic and nonrelativistic Hopf-algebraic counterparts have
been proposed in  \cite{chi} and \cite{daszkiewicz} respectively.\\
The second kind of mentioned deformations  introduces the
Lie-algebraic type of space-time noncommutativity
\begin{equation}
[\;{\hat x}_{\mu},{\hat x}_{\nu}\;] = i\theta_{\mu\nu}^{\rho}{\hat
x}_{\rho}\;, \label{noncomm1}
\end{equation}
with particularly chosen coefficients $\theta_{\mu\nu}^{\rho}$ being
constants. The corresponding Poincare quantum groups have been
introduced in \cite{kappaP}-\cite{lie2}, while the suitable Galilei
algebras  - in \cite{kappaG} and \cite{daszkiewicz}.\\
The last kind of quantum space, so-called quadratic type of
noncommutativity
\begin{equation}
[\;{\hat x}_{\mu},{\hat x}_{\nu}\;] =
i\theta_{\mu\nu}^{\rho\tau}{\hat x}_{\rho}{\hat
x}_{\tau}\;\;\;;\;\;\;\theta_{\mu\nu}^{\rho\tau} = {\rm const.}\;,
\label{noncomm2}
\end{equation}
has been  proposed  in \cite{qdef}, \cite{paolo} and \cite{lie2} at
relativistic and in \cite{daszkiewicz2} at nonrelativistic level.

Recently, in the series of papers \cite{Daszkiewicz:2008fn}-\cite{Harikumar:2010aw}, there has been demonstrated that the different (mentioned above)
types of space-time noncommutativity produce in
particle models  the additional dynamical terms. Particulary, in  article \cite{Daszkiewicz:2008fn} it has been shown, that when two spatial coordinates commute to time in Lie-algebraically way, there is generated the additional acceleration of nonrelativistic  particle
moving in  constant external field force. Similar investigations for quantum space with two spatial directions commuting to the proper function of classical
time have been performed for  particle in constant  force \cite{Daszkiewicz:2011jq} as well as for  oscillator model \cite{Daszkiewicz:2008ij}. Besides, the other types of space-time noncommutativity (with two spatial  directions commuting to space in Lie-algebraically and quadratically way) have been studied in  \cite{Daszkiewicz:2008fn}  and in article \cite{Miao:2009jw}. It should be noted, however, that the most interesting results
have been obtained in  paper \cite{Harikumar:2010aw}, in which the authors considered nonrelativistic particle moving in gravitational potential on the well-known
$\kappa$-Galilei space-time \cite{kappaG}.  Particulary, it has been demonstrated that the generated by such a type of space-time noncommutativity additional force term can be
identified with so-called Pioneer anomaly  \cite{Anderson:1998jd}. The comparison of obtained result with the proper observational data
\cite{Anderson:1998jd}, \cite{Anderson:2001sg}  permitted  to fix the deformation
parameter $\kappa$.

In this article we generate the (additional) time-dependent force terms in Newton equation of particle moving in constant force $\vec{F}$ in the completely new way, i.e. by the   transformation of classical space of the form
\begin{eqnarray}
&x_i& \longrightarrow \;\;x_i+ a_i(t)\;,\label{trans1}\\
&t& \longrightarrow \;\;t\;. \label{trans2}
\end{eqnarray}
Formally, the rules (\ref{trans1}) and (\ref{trans2}) can be realized in the framework of so-called doubly enlarged Newton-Hooke (undeformed) quantum group $\;{\mathcal U}_{0}(\widehat{\widehat{NH}}_{+})$, for which \cite{Gomis:2008jc}\footnote{In \cite{Gomis:2008jc} there is considered only the hyperbolic
 (De-Sitter) case. However, the trigonometric (anti-De-Sitter) transformation can be easily
 obtained by changing cosmological constant  $\tau$ into  $i \tau$ in all above equations.}, \footnote{Here, we take under consideration the most general (known) type of
 nonrelativistic transformations of classical space.}
\begin{eqnarray}
a_i(t) &=& a_i \cosh \left(\frac{t}{\tau}\right) + v_i  \tau \sinh \left( \frac{t}{\tau}\right) + 2 b_i  \tau^2
\left(\cosh \left(\frac{t}{\tau}\right)  - 1\right) + \nonumber\\
&+&6c_i\tau^3\left(\sinh \left(\frac{t}{\tau}\right)  - \frac{t}{\tau}\right)\;,\label{realization}
\end{eqnarray}
and for which, in $\tau$ approaching infinity limit, we have\footnote{The formula (\ref{realizationlimit}) defines the transformation rules  for so-called doubly enlarged Galilei Hopf structure $\;{\mathcal U}_{0}(\widehat{\widehat{G}})$.}
\begin{eqnarray}
a_i(t) &=& a_i + v_i  t +  b_i  t^2+ c_it^3\;.\label{realizationlimit}
\end{eqnarray}
It should be also noted, that by the proper contraction schemes ($\tau \to \infty$) one can get from $\;{\mathcal U}_{0}(\widehat{\widehat{NH}}_{+})$ the other nonrelativistic symmetry
groups, such as: acceleration-enlarged Newton-Hooke, acceleration-enlarged Galilei and Galilei classical Hopf algebras, respectively.

In the  next step of our investigations, we compare the obtained result with the mentioned above model  of nonrelativistic particle defined
 on the  quantum space of the form\footnote{We consider particle moving in the presence of external constant field force $\vec{F}$.}
\begin{equation}
[\;t,\bar{x}_{i}\;] =[\;\bar{x}_{1},\bar{x}_{3}\;]  = [\;\bar{x}_{2},\bar{x}_{3}\;]  =
0\;\;\;,\;\;\; [\;\bar{x}_{1},\bar{x}_{2}\;]  =
if({t})\;\;;\;\;i=1,2,3
\;, \label{spaces}
\end{equation}
with
\begin{eqnarray}
f({t})&=&
f_{\kappa_1}\left(\frac{t}{\tau}\right) = \kappa_1\,\cosh^2
\left(\frac{t}{\tau}\right)\;, \label{w2}\\
f({t})&=&
f_{\kappa_2}\left(\frac{t}{\tau}\right) =\kappa_2\tau\, \cosh
\left(\frac{t}{\tau}\right)\sinh \left(\frac{t}{\tau}\right) \;,
\label{w3}
\end{eqnarray}
\begin{eqnarray}
f({t})&=&
f_{\kappa_3}\left(\frac{t}{\tau}\right) =\kappa_3\tau^2\,
\sinh^2 \left(\frac{t}{\tau}\right) \;, \label{w4}\\
f({t})&=&
 f_{\kappa_4}\left(\frac{t}{\tau}\right) = 4\kappa_4
 \tau^4\left(\cosh\left(\frac{t}{\tau}\right)
-1\right)^2 \;, \label{w5}\\
f({t})&=&
f_{\kappa_5}\left(\frac{t}{\tau}\right) =  \kappa_5\tau^2
\left(\cosh\left(\frac{t}{\tau}\right)
-1\right)\cosh \left(\frac{t}{\tau}\right)\;, \label{w6}\\
f({t})&=&
f_{\kappa_6}\left(\frac{t}{\tau}\right) = \kappa_6\tau^3
\left(\cosh\left(\frac{t}{\tau}\right) -1\right)\sinh
\left(\frac{t}{\tau}\right)\;,\label{w7}
\end{eqnarray}
and (for $\tau \to \infty$)
\begin{eqnarray}
f(t) &=& f_{\kappa_1}({t}) = \kappa_1\;,\label{ggnw2}\\
f(t) &=& f_{\kappa_2}({t}) = \kappa_2\,t\;,\label{ggnw3}\\
f(t) &=& f_{\kappa_3}({t}) = \kappa_3\,t^2\;,\label{ggnw4}\\
f(t) &=& f_{\kappa_4}({t}) = \kappa_4\,t^4\;,  \label{ggnw5}\\
f(t) &=& f_{\kappa_5}({t}) = \frac{1}{2}\kappa_5\,t^2\;, \label{ggnw6}\\
f(t) &=& f_{\kappa_6}({t}) = \frac{1}{2}\kappa_6\,t^3\;. \label{ggnw7}
\end{eqnarray}
The commutation relations (\ref{spaces}) has been provided in the framework of twist procedure of acceleration-enlarged Newton-Hooke
Hopf algebra $\;{\mathcal U}_{0}({\widehat{NH}}_{+})$ \cite{Daszkiewicz:2010bp}\footnote{In this article we consider the most general (known)
twist deformation of nonrelativistic symmetries.}. Here, by the comparison of  both treatments, we find the direct link
between transformation functions $a_i(t)$ and  time-dependent noncommutativity $f(t)$.

The paper is organized as follows. In second section  we consider 
the noncommutative model of nonrelativistic particle moving in constant field force $\vec{F}$. 
Further, we provide its commutative counterpart which incorporates the transformation rules (\ref{trans1}) and  (\ref{trans2}). In section three we compare the both (described above) force term generating  procedures. The final remarks are mentioned in the last section.

\section{{{Generating of the additional force terms in Newton equation}}}

\subsection{{{Generating by the space-time noncommutativity (\ref{spaces}) - the first treatment}}}

Let us now turn to the  dynamical models in which the additional force terms are generated by  space-time noncommutativity. Firstly, we start with the
following phase space\footnote{We use the correspondence relation $\{\;a,b\;\}
= \frac{1}{i}[\;\hat{a},\hat{b}\;]$  $(\hbar = 1)$.}
\begin{equation}
\{\;t,{ {\bar x}}_{i}\;\} = 0 \;\;\;,\;\;\;\{\;{ {\bar x}}_{1},{ {\bar
x}}_{2}\;\}  = f(t)\;\;\;,\;\;\;
\{\;{ {\bar x}}_{1},{{\bar
x}}_{3}\;\}= 0=\{\;{ {\bar x}}_{2},{ {\bar
x}}_{3}\;\} \;, \label{beyond}
\end{equation}
\begin{equation}
\{\;{ {\bar x}}_{i},{\bar p}_j\;\} = \delta_{ij}\;\;\;,\;\;\;\{\;{
{\bar p}}_{i},{ {\bar p}}_{j}\;\} = 0\;, \label{genin2a}
\end{equation}
corresponding to the commutation relations (\ref{spaces}). One can check that the  relations (\ref{beyond}),
(\ref{genin2a}) satisfy the Jacobi identity and for deformation
parameters $\kappa_a$ running to zero become classical. Next, we define
the Hamiltonian function for
nonrelativistic particle moving in constant field force $\vec{F}$ as follows 
\begin{eqnarray}
H(\bar{p},\bar{x}) = \frac{1}{2m}\left({\bar p}_{1}^2 +
{\bar p}_{2}^2 + {\bar p}_{3}^2\right) -  \sum_{i=1}^{3}F_i \bar{x}_i\;.\label{ham1}
\end{eqnarray}
In order to analyze the above system we represent the
noncommutative variables $({\bar x}_i, {\bar p}_i)$ on classical
phase space $({ x}_i, { p}_i)$ as  (see e.g. \cite{lumom}-\cite{giri})
\begin{equation}
{\bar x}_{1} = { x}_{1} - \frac{f_{\kappa_a}(t)}{2}
p_2\;\;\;,\;\;\;{\bar x}_{2} = { x}_{2} +\frac{f_{\kappa_a}(t)}{2}
p_1\;\;\;,\;\;\; {\bar x}_{3}= x_3 \;\;\;,\;\;\; {\bar p}_{i}=
p_i\;, \label{rep}
\end{equation}
where
\begin{equation}
\{\;x_i,x_j\;\} = 0 =\{\;p_i,p_j\;\}\;\;\;,\;\;\; \{\;x_i,p_j\;\}
=\delta_{ij}\;. \label{classpoisson}
\end{equation}
Then, the  Hamiltonian (\ref{ham1})  takes the form
\begin{eqnarray}
{{H}}({ p},{ x})=H_f(t) = \frac{1}{2m}\left({ p}_{1}^2 +
{ p}_{2}^2 + { p}_{3}^2\right) - \sum_{i=1}^{3}F_i x_i + F_1 \frac{f_{\kappa_a}(t)}{2}p_2 -
F_2 \frac{f_{\kappa_a}(t)}{2}p_1 \label{hamoscnew}\;.
\end{eqnarray}
Using the formulas (\ref{classpoisson}) and (\ref{hamoscnew}) one gets
the following canonical Hamiltonian equations of motions $(\dot{{o}}_i = \frac{d}{dt}o_i = \{\;o_i,H\;\})$
\begin{eqnarray}
&&\dot{x}_{1} = \frac{p_1}{m} - \frac{f_{\kappa_a}(t)
}{2}F_2 \;\;\;,\;\;\; \dot{p}_{1} = F_1
\;,\label{ham1a}\\
 &~~&~\cr
&&\dot{x}_{2} = \frac{p_2}{m} + \frac{f_{\kappa_a}(t)
}{2}F_1\;\;\;,\;\;\; \dot{p}_{2} = F_2
\;,\label{ham2a}\\
&~~&~\cr
&&~~~~~\dot{x}_{3} = \frac{p_3}{m}\;\;\;,\;\;\;\dot{p}_{3} =
F_3\;,\label{ham3a}
\end{eqnarray}
which when combined yield the proper Newton law
\begin{equation}
\left\{\begin{array}{rcl} m\ddot{x}_1  &=&{F_1} - \frac{m\dot{f}_{\kappa_a}(t)
}{2}F_2 = G_1(t)\\
 &~~&~\cr
 m\ddot{x}_2  &=&
{F_2} + \frac{m\dot{f}_{\kappa_a}(t)
}{2}F_1 = G_2(t)\\
 &~~&~\cr
 m\ddot{x}_3  &=& {F_3} = G_3
 \;.\end{array}\right.\label{dddmixednewton1}
\end{equation}

Firstly, by trivial integration one can find the solution of  above system; it
 looks as follows
 \begin{eqnarray}
 \left\{\begin{array}{rcl}{x}_{1}(t) &=& \frac{F_1}{2m}t^2  + v^0_1t - \frac{F_2}{2}\int_{0}^{t}{f}_{\kappa_a}(t')dt'\\
 &~~&~\cr
 {x}_{2}(t) &=& \frac{F_2}{2m}t^2  + v^0_2t + \frac{F_1}{2}\int_{0}^{t}{f}_{\kappa_a}(t')dt'\\
 &~~&~\cr
{x}_{3}(t) &=& \frac{1}{2m}F_3t^2  + v^0_2t +x^0_3\;,\end{array}\right.\label{sol3}
\end{eqnarray}
with $v^0_i$ and $x_3^0$ denoting the initial velocity and position of particle respectively.
Next, one should  observe that the noncommutativity (\ref{spaces}) generates the new, time-dependent force term
$\vec{G}(t) = \left[\;G_1(t),G_2(t),G_3\;\right]$, which for deformation parameters $\kappa_a$ approaching
zero reproduces undeformed force $\vec{F}$. 
Finally, it should be  noted that for $f(t) = \lim_{\tau \to \infty}f_{\kappa_1}(t)= \kappa_1 = \theta$
and $f(t) = \lim_{\tau \to \infty}f_{\kappa_2}(t)= \kappa_2 t$
(see formulas (\ref{ggnw2}) and (\ref{ggnw3}) respectively)
we recover two models provided in \cite{Daszkiewicz:2008fn}. First of them does not introduce any modification of Newton equation, while
the second one generates the constant acceleration of particle. 

\subsection{{{Generating by the transformation of classical space (\ref{trans1})-(\ref{realization}) - the second treatment}}}

Let us now turn to the second model in which the force terms are generated by the transformation of classical space (\ref{realization}). Firstly, we start
with the following equation of motion
\begin{eqnarray}
\left\{\begin{array}{rcl} m\ddot{x}_1  &=&{F_1} \\
 &~~&~\cr
 m\ddot{x}_2  &=&
{F_2} \\
 &~~&~\cr
 m\ddot{x}_3  &=& {F_3}
 \;,\end{array}\right.\label{start}
\end{eqnarray}
defined on the commutative (standard) space-time.
Next, by
using transformation rules (\ref{trans1}) with function $a_3(t)$ equal zero, we get the following Newton law in the nonrelativistic space-time with
changed space coordinates (see (\ref{trans1}))
\begin{eqnarray}
\left\{\begin{array}{rcl} m\ddot{x}_1  &=&{F_1} + m{\ddot a}_1(t) = H_1(t)\\
 &~~&~\cr
 m\ddot{x}_2  &=&
{F_2} + m\ddot{a}_2(t) = H_2(t)\\
 &~~&~\cr
 m\ddot{x}_3  &=& {F_3} = H_3
 \;,\end{array}\right.\label{secondnewton1}
\end{eqnarray}
i.e. there appeared in  Newton equation the additional force term given by the function ${\ddot a}_i(t)$. Moreover,
let us notice that only for Galileian transformation
\begin{eqnarray}
a_i(t) = a_i + v_i t\;,
\end{eqnarray}
the equation of motion (\ref{start}) remains unchanged. Besides, one should  observe that the solution of (\ref{secondnewton1}) is given by
 \begin{eqnarray}
\left\{\begin{array}{rcl} {x}_{1}(t) &=& \frac{F_1}{2m}t^2  + v^0_1t + x^0_1 + a_1(t)\\
 &~~&~\cr
{x}_{2}(t) &=& \frac{F_2}{2m}t^2  + v^0_2t + x^0_2 +a_2(t)\\
 &~~&~\cr
{x}_{3}(t) &=& \frac{1}{2m}F_3t^2  + v^0_2t +x^0_3\;,\end{array}\right.
\label{secondsol3}
\end{eqnarray}
and (for example) for function (\ref{realizationlimit}), it takes the form
\begin{eqnarray}
\left\{\begin{array}{rcl} {x}_{1}(t) &=& c_1t^3+\left(\frac{F_1}{2m}+ b_1\right)t^2  + \left(v^0_1+v_1\right)t + x^0_1+a_1   \\
 &~~&~\cr
{x}_{2}(t) &=& c_2t^3+\left(\frac{F_2}{2m}+b_2\right)t^2  + \left(v^0_2+v_2\right)t + x^0_2+a_2 \\
 &~~&~\cr
{x}_{3}(t) &=& \frac{1}{2m}F_3t^2  + v^0_2t +x^0_3\;.\end{array}\right.
\label{secondsolform}
\end{eqnarray}
Consequently, we see that as in the pervious treatment there appeared in model the new,
time-dependent force $\vec{H}(t) = \left[\;H_1(t),H_2(t),H_3\;\right]$
defined by the equation (\ref{secondnewton1}). 

\section{Comparison  of the both approaches}

Let us now compare the formulated above treatments. 
Firstly, one can observe that the new, time-dependent forces $\vec{G}(t)$ and $\vec{H}(t)$ are exactly the same when
\begin{eqnarray}
\ddot{a}_1(t) = - \frac{\dot{f}_{\kappa_a}(t)
}{2}F_2\;\;\;,\;\;\;
\ddot{a}_2(t) =  \frac{\dot{f}_{\kappa_a}(t)
}{2}F_1 \;.\label{triondnewpot1}
\end{eqnarray}
Unfortunately, such a situation appears only for (see (\ref{realizationlimit}) and (\ref{ggnw3}))\footnote{The equality of force terms in both
treatments appears only for cosmological constant $\tau$ approaching infinity.}
\begin{eqnarray}
f_{\kappa_2}(t) &=& \kappa_2 t\;\;\;,\;\;\; a_{1}(t) = a_1 + v_1t + b_1t^2 \;\;\;;\;\;\; b_1 = - \frac{\kappa_2}{4}F_2\;,\label{codi1a}\\
a_{2}(t) &=& a_2 + v_2t + b_2t^2 \;\;\;;\;\;\; b_2 = \frac{\kappa_2}{4}F_1\;,\label{codi1b}
\end{eqnarray}
as well as for (see formulas (\ref{realizationlimit}), (\ref{ggnw4}) and (\ref{ggnw6}))
\begin{eqnarray}
a_1(t) &=& a_1 + v_1t + c_1t^3 \;\;\;;\;\;\;c_1 = - \frac{\kappa}{6}F_2\;,\label{codi2a}\\
a_2(t) &=& a_2 + v_2t + c_2t^3 \;\;\;;\;\;\; c_2 =  \frac{\kappa}{6}F_1\;,\label{codi2b}\\
f_{\kappa_3}(t) &=& \kappa_3t^2 \;=\; f_{\kappa_5}(t) \;=\; \frac{1}{2}\kappa_5 t^2 \;=\; \kappa t^2\;.\label{codi2}
\end{eqnarray}
Then, for the choices (\ref{codi1a}), (\ref{codi1b}) and (\ref{codi2a})-(\ref{codi2}), we get the following equalities
\begin{eqnarray}
H_1(t) &=& G_{1}(t) \;=\; F_1-\frac{m\kappa_2}{2}F_2\;,\label{equal1a}\\
H_2(t) &=& G_{2}(t) \;=\; F_2+\frac{m\kappa_2}{2}F_1\;,\label{equal1b}\\
H_3 &=& G_{3} \;=\; F_3\;,
\label{equal1c}
\end{eqnarray}
and
\begin{eqnarray}
H_1(t) &=& G_{1}(t) \;=\; F_1-m{\kappa}F_2t\;,\label{equal2a}\\
H_2(t) &=& G_{2}(t) \;=\; F_2+m{\kappa}F_1t\;,\label{equal2b}\\
H_3 &=& G_{3} \;=\; F_3\;,\label{equal2c}
\end{eqnarray}
respectively.
Besides, it seems that the second treatment (by the classical groups) has one advantage - it follows from
(\ref{dddmixednewton1}) and (\ref{secondnewton1}) that the second approach (contrary to the first one) does not need the presence
of initial constant force $\vec{F} = \left[\;F_1,F_2,F_3\;\right]$ to generate the additional dynamical terms. In fact, when
one puts $\vec{F} = \left[\;0,0,0\;\right]$ in the equation of motion (\ref{dddmixednewton1}) then its right side vanishes, while  in the equation (\ref{secondnewton1}) there are still presented the additional forces of the form
\begin{eqnarray}
H_1 (t) = m\ddot{a}_1(t)\;\;\;,\;\;\;H_2 (t) = m\ddot{a}_2(t)\;.\label{still}
\end{eqnarray}

\section{Final remarks}

In this article we compare two ways of force term generating in the model of nonrelativistic particle moving in
constant external field force. First of them uses the space-time noncommutativity (\ref{spaces}) while the second one is based on the transformation
rules of classical space (\ref{trans1}). Particulary, we find for which functions $f(t)$ and $a_i(t)$ the generated force terms
are the same in both treatments. Finally, it should be noted that performed in this article considerations concern only the simplest model of nonrelativistic particle moving in constat force $\vec{F}$. However, they can be extended to the arbitrary norelativistic system and then, the necessary calculations become much more complicated but the general mechanism remains the same.

\section*{Acknowledgments}
The author would like to thank J. Lukierski
for valuable discussions. This paper has been financially  supported  by Polish
NCN grant No 2011/01/B/ST2/03354.

\end{document}